# Non-Stokesian dynamics of magnetic helical nanoswimmers under confinement

Alireza Fazeli,[1,2] Vaibhav Thakore,[1,2] Tapio Ala-Nissila,[3,4] Mikko Karttunen[2,5,6*]

[1]Department of Mathematics, Western University, 1151 Richmond Street, London, Ontario, N6A 5B7, Canada.
[2]Center for Advanced Materials and Biomaterials Research, Western University, 1151 Richmond Street, London, Ontario, N6A 5B7, Canada.
[3]Multiscale Statistical and Quantum Physics Group, Quantum Technology Finland Centre of Excellence, Department of Applied Physics, Aalto University, P.O. Box 11000, FI-00076 Aalto, Espoo, Finland.
[4]Interdisciplinary Centre for Mathematical Modelling, Department of Mathematical Sciences, Loughborough University, Loughborough, Leicestershire, LE11 3TU, United Kingdom.
[5]Department of Physics and Astronomy, Western University, 1151 Richmond Street, London, Ontario, N6A 5B7, Canada.
[6]Department of Chemistry, Western University, 1151 Richmond Street, London, Ontario, N6A 3K7, Canada.

[*]mkarttu@uwo.ca

**Abstract:** Electromagnetically propelled helical nanoswimmers offer great potential for nanorobotic applications. Here, the effect of confinement on their propulsion is characterized using lattice-Boltzmann simulations. Two principal mechanisms give rise to their forward motion under confinement: 1) pure swimming, and 2) the thrust created by the differential pressure due to confinement. Under strong confinement, they face greater rotational drag, but display a faster propulsion for fixed driving frequency in agreement with experimental findings. This is due to the increased differential pressure created by the boundary walls when they are sufficiently close to each other and the particle. Two new analytical relations are presented: 1) for predicting the swimming speed of an unconfined particle as a function of its angular speed and geometrical properties, and 2) an empirical expression to accurately predict the propulsion speed of a confined swimmer as a function of the degree of confinement and its unconfined swimming speed. At low driving frequencies and degrees of confinement, the systems retain the expected linear behavior consistent with the predictions of the Stokes equation. However, as the driving frequency and/or the degree of confinement increase, their impact on propulsion leads to increasing deviations from the Stokesian regime and emergence of nonlinear behavior.

# INTRODUCTION

Life in the Stokesian regime is largely dominated by viscous forces where advective inertial forces play no role, *i.e.*, the present is determined entirely by the instantaneous forces and the system possesses no memory of its past (*1*). Due to the time reversibility of the Stokes equations, locomotion in the microscopic scale is subject to the scallop theorem: in viscous-dominated Newtonian fluids, for a swimming gait to give rise to motion, it must not consist of a time-reversible series of body deformations (*2*, *3*). Therefore, non-inertial swimming microorganisms have evolved to use other methods to create forward motion, such as viscous drag anisotropy (*4*). One well-known and widely studied example is the flagellar swimming of *E. coli*, which has a single anterior flagellum, a rotary joint, and a motor to rotate the flagellum (*5*, *6*). Subsequent and continuous deformation of the flagellum is characterized by a generalized helical motion in three dimensions (3D) that lacks viscous drag symmetry (*7*), enabling the bacterium to propel itself forward.

Understanding the hydrodynamics of biomimetic micro- and nanoswimmers in constrained spaces is critical to address various environmental and medical challenges. For instance, bacteria play a crucial role in purifying aquifers, particularly when there is adequate groundwater flow that prevents motile bacteria from adhering to soil particles (*8*). As another example, the motility of spirochetes enables them to penetrate tissues more effectively compared to their non-motile counterparts (*9*). Ongoing clinical and theoretical research is leveraging such insights to explore the potential of these small-scale robots to transport drugs to specific areas within the body (*10*–*12*). As a result, the interest in creating biomimetic swimmers at the nano- and micro-scales, such as asymmetric colloidal particles that are gravitactic, similar to paramecium, or rheotactic, akin to sperm and bacteria, has grown steadily over the past decade (*13*–*17*). Their applications include tasks such as carrying cargo in microfluidic chips (*18*), transporting drugs and genetic material into cells and tissues (*11*, *19*–*21*), fluid mixing to speed up chemical reactions (*22*), and motorizing sperm cells (*23*). In many cases, accurate and rapid directed mobility of the swimmer under strong confinement is a key requirement.

Using a combination of experimental and computational methods, Jung *et al*. towed superimposed copper helices (*i.e.*, of opposite chirality as inspired by the geometry of spirochaetes) in a Stokes fluid and reported that the effect of the confinement was negligible on their propulsion purportedly due to the small ratio of the body size to the container width (*24*). Spagnolie and Lauga, by means of simulations and far-field approximations, concluded that the boundary interactions become significant when the distance from the swimmer to the wall is comparable to the swimmer's length (*25*). Using the boundary element method (BEM), Liu *et al*. investigated the microscale propulsion of a helical flagellum rotating in a circular tube. They demonstrated that at fixed angular speed propulsion grows monotonically as the confinement becomes tighter (*26*). Vizsnyiczai *et al*. performed experiments and reported that, in contrast with theoretical predictions, the average speed of *E. coli* increased when moving through a sequence of rectangular microtunnels with decreasing widths until it reached a maximum value. They also showed that strong confinement results in a transition to one-dimensional swimming where bacteria only explore the neighborhood of the capillary axis and move with a speed that is higher than in lesser degree of confinement (*27*). However, so far there has been no systematic effort to understand and explain the underlying causes of these, often contrasting (*21*, *28*), experimental and computational results.

The commonly employed analytical and computational methods for modeling the swimming of rigid helices and flagellar propulsion are the resistive force theory (RFT) (*29*), slender-body theory (SBT) (*30, 31*), boundary integral formulation (BIF) (*32*) and its expansions, *e.g.*, BEM (*33–35*). The RFT assumes that the segments of the helical body do not create fluid disturbances or influence the motions of other segments. Hydrodynamic forces on an infinitesimal segment are determined by the force (or drag) coefficient proportional to the local velocity, and propulsive velocity calculations involving the summation of the drag and thrust contributions of the rotating helical particles (*29, 36*). This assumption is, however, inconsistent with the true hydrodynamic situation in which viscous effects dominate and produce long-range hydrodynamic interactions (*37*). There have been reports on qualitative and quantitative differences between the predictions of RFT and experimental results (*38*) particularly for the cases where the helical loops are more closely packed at higher pitch angles (*39*). In contrast, the SBT incorporates these interactions by considering fluid response resulting from the moving segments of the helical body (*37*). This is achieved by assuming a local point force represented by Stokeslets, uniformly distributed along the centerline of the helical body (*30, 31*). The regularized SBT extends this concept by using a shape function that distributes the force in a small blob around a point, rather than assuming a Dirac delta function (*40*). Findings from both the SBT and regularized SBT exhibit resemblances in relation to flagellar propulsion of microorganisms (*38*). However, all these studies either use the linearized Stokes approximation for modeling the fluid-swimmer interaction dynamics or crucially rely on it for the interpretation and analysis of their experimental observations. Our results here show that the linear Stokes approximation is not always applicable or valid at propulsion frequencies typically used for driving and steering magnetic helical nanoswimmers under nanoscale confinement.

Recently, the Lattice-Boltzmann Method (LBM) combined with explicit particle-level molecular dynamics simulations has been employed to model helical and flagellar swimming (*41–45*). For a review of LBM, see, *e.g.*, Ref. (*46*). LBM is a computationally efficient mesoscale simulation technique based on the Boltzmann transport equation that is valid at relatively large Knudsen numbers (*47, 48*). LBM takes full-scale nonlinear hydrodynamics including inertial and viscous forces into account and is well suited for the task (*48*). This work quantifies and formulates the nonlinear hydrodynamic effects of confinement on the propulsion characteristics of magnetically driven helical filaments with computer simulations using LBM to systematically understand and explain the underlying physics of the aforementioned experimental and computational observations (*21, 24–28*).

In this work, the swimmers are subjected to different degrees of confinement and driven using a rotating magnetic field. The degree of confinement is defined as $\Gamma = D_o/L_z$, where $D_o$ and $L_z$ are the helix's outer diameter and the distance between the upper and lower boundary walls (*i.e.*, *z*-walls), respectively. Unexpectedly, local nonlinear dynamics are observed even for driving frequencies $f$ well below the step-out frequencies $f_{so}$ of the systems (*i.e.*, the frequency requiring the entire available magnetic torque to maintain synchronous rotation); steady and linear dynamics is recovered over longer time scales for $f < f_{so}$, consistent with the Stokes equations. However, for frequencies close to and above the step-out frequency, nonlinearities start to dominate, and the propulsion speed can no longer be described as a linear function of the angular speed. Under higher degrees of confinement, nonlinearities are present independent of $f$, and a transition to one-dimensional-like swimming emerges for a critical degree of confinement, *i.e.*, $\Gamma = 0.33$. Furthermore, a temporal analysis of the swimming dynamics and trajectories analyzed as a function of the degree of confinement showed that as the fluidic channel narrows, the swimmers encounter greater rotational drag, but paradoxically, achieve a faster propulsion. A decomposition of the total propulsion into pure swimming and thrust from the confined fluid shows that the extra propulsion results from the differential pressure created by the side walls and overcomes the increased rotational drag. The effect of confinement is observed to diminish with increasing channel width and nearly vanish when the width exceeds four times the hydrodynamic diameter of the swimmer.

In the next section, we first explain the step-out regime and examine the impact of the step-out frequency on the angular speed of an individual nanoswimmer. A comparison of the velocities in the frequency domain is then presented to understand the complex relationships between the degree of confinement, viscous drag, and differential pressure (fore and aft) on the dynamics of a helical nanoparticle. This is followed by the introduction of a new closed-form equation to establish a direct linear relationship between the translational and angular velocities of the helical filament. After that, we propose an empirical equation to accurately predict the propulsion speed of a confined swimmer as a function of an unbounded particle's propulsion speed and the ratio between the hydrodynamic characteristic length scale linked to the confined flow and that of the swimmer. The proposed equation is expected to serve as an important heuristic tool in the design of systems utilizing swimmers in applications involving transport in constricted fluidic environments such as tissues, veins, and lab-on-a-chip micro/nanofluidic devices. The key findings are summarized in Discussion followed by the Methods section, which outlines the computational model employed in this study. It describes the essential model parameters and geometric properties of the particles considered in the simulations using LBM. Incorporation of magnetic interactions within the system using the resistance matrix formulation is also described to explain the basic analytical framework for swimming solely based on viscous drag asymmetry.

## RESULTS

### Understanding the step-out regime

Due to the magnetic interaction, the angular speed $(\Omega)$ of a magnetic helical particle increases linearly with $f$ and so does its translational speed $(U)$. This linearly increasing relation holds valid up to field frequencies close to the step-out frequency (*43, 49–51*) given by $f_{so} = MB/K_{\parallel}$ (*43*), where $K_{\parallel}$ is the rotational drag coefficient. Upon approaching $f_{so}$, irregular episodes of phase slip between the rotations of the field and the swimmer emerge, *i.e.*, the blue stripes in [Fig. 1](). This occurs because the applied magnetic torque is no longer sufficient to effectively overcome the increased viscous drag torque; thus, some of the rotational cycles of the swimmer are damped prematurely. Therefore, the magnetic field fails to keep the rotation of the swimmer steady, and the particle starts to lag the magnetic field intermittently. This leads to a stark decline in the average angular speed, and therefore, the average propulsion speed of the helix. A further increase in the field frequency causes the phase slips to become more frequent and irregular as more rotation cycles are damped out thereby leading to a further reduction in the average propulsion speed of the swimmer. The same observation has been reported numerically and experimentally, for example, in Refs. (*42, 50, 52, 53*).

It is important to note that while the phase flips observed here (*i.e.*, the moments when the direction of rotation of the helix is reversed momentarily) are reminiscent of the characteristics of a system running at (damped) natural frequency, by writing the equation of motion using D'Alembert's principle (assuming infinite stiffness for the submerged helix), it can be shown that these systems are indeed overdamped and their step-out frequencies are significantly lower than their natural frequencies. The phase flips occur due to the changes in frequency resulting from the competition among the two opposing torque vectors in the step-out regime.

The results on the average angular speeds of the helix for field frequencies smaller and greater than $f_{so}$ are reported in Fig. S1 as a function of the degree of confinement and viscosity $(\mu_d)$. They illustrate that for the same field frequency, the particle rotates faster when $\Gamma$ is smaller. Regardless of the value of $\Gamma$, the average angular speed drops sharply when the field frequency exceeds $f_{so}$. Higher viscosities lead to a smaller $f_{so}$ and in all cases, increasing $\Gamma$ results in an earlier onset for the phase slips, leading to even lower step-out frequencies. Importantly, this suggests that the helical filament experiences a greater rotational drag when swimming in narrow channels with higher degrees of confinement.

**Impact of confinement on the particle dynamics**
In the Stokesian regime, a helical filament is predicted to travel with some degree of conical wobble in addition to its translation if it has a non-integer number of turns (*54*). Swimmer H0 (see Table S1 for its properties) has 1.25 turns and therefore its trajectories in the *y-z* plane are expected to be elliptical in shape. It is seen in Fig. 2 that this holds true except for cases with $\Gamma > 0.25$. As $\Gamma$ increases, the proximity of the swimmer to the boundary walls generates greater feedback (see also Fig. 3) thereby giving rise to a stronger boundary effect and the emergence of nonlinearities. This disrupts the expected *y-z* trajectory of the swimmer and amplifies drag, leading to an even greater feedback when the field frequency is increased under stronger confinement (*e.g.*, $\Gamma = 0.33$ Fig. 2C). The directionality remains independent of confinement, in agreement with the analytical results obtained using BIF in Ref. (*55*) but only for $\Gamma < 0.50$, wherein the net forces in the *y*- and *z*-directions are nearly zero. In contrast, for the case of $\Gamma = 0.50$ negative tangential force due to traction force pushes the swimmer sideways in a direction opposite to its chirality during its forward motion. This is aligned with the conclusions of Ref. (*35*) in which Shum *et al.*, using BEM, reported that the configuration in the center of the channel and parallel to the walls was a stable position for their model bacterium shape (*i.e.*, a helix with an attached ellipsoid/cargo) when, in our nomenclature, $\Gamma = 0.25$–$0.33$.

There are analytical, *e.g.*, (*35*), and experimental, *e.g.*, (*27*), reports on a transition to one-dimensional-like swimming under strong confinement characterized by a situation where the range of changes in the *y*- and *z*-displacements of the swimmer are reduced significantly. Our results show that this occurs at a critical $\Gamma$, *i.e.*, $\Gamma = 0.33$ in Fig. 2C, and only when $f < f_{so}$. For $\Gamma$ greater than this critical value, the swimmer enters a confinement-induced unstable equilibrium state wherein the feedback from the side walls shifts the direction of swimming, causing the swimmer to drift sideways in a direction opposite to its chirality during its forward motion (see Fig. 2D).

Ghosh *et al.* studied changes in velocity in helical propulsion (*55*). Based on their experiments, they reported that as opposed to bulk fluid, the velocity power spectra for propellers under confinement were not perfectly flat and showed peaks at the multiples of the driving frequency. They proposed detailed numerical simulations to be performed to understand these effects.

The Discrete Fourier Transform (DFT) is a technique used to analyze and transform a discrete signal from the time domain to the frequency domain. It decomposes a signal into its constituent sinusoidal components of different frequencies. DFT analysis is performed by applying a DFT algorithm such as the Fast Fourier Transform (FFT) (*57*). The algorithm calculates the complex amplitudes of the sinusoidal components at different frequencies present in the signal. These complex amplitudes represent the magnitude and phase information of each frequency component. In a DFT analysis, nonlinearities can introduce additional frequencies that are integer multiples of the fundamental frequency referred to as harmonics in the frequency spectrum.

Here, an FFT analysis of the velocity in the direction perpendicular to the *z*-walls reveals the presence of higher order harmonics (Fig. 4) in agreement with Ref. (*55*). Harmonics can arise from factors including both linear and nonlinear phenomena. In a linear system, harmonics can be present due to periodic non-sinusoidal signals or, equivalently, due to interference arising from the presence of multiple frequencies in the input signal. Nonetheless, in nonlinear systems, harmonics are expected to be more pronounced and exhibit characteristics such as distortion and amplitude modulation. In this context, it is worth noting that all simulations in this study employ a single frequency harmonic driving force using an external rotating magnetic field and, consequently, any higher order harmonics observed can be attributed to interference effects and system nonlinearities alone. For $\Gamma \leq 0.25$ high order harmonics are present only if the driving frequency is equal to or greater than the step-out frequency. In contrast, for $\Gamma > 0.25$, nonlinear harmonics are present even for the smallest driving frequency examined, *i.e.*, 300 kHz, denoted by the arrows in Fig. 4C and D. This implies that at high degrees of confinement and/or driving frequencies, nonlinearities are more pronounced, leading to a deviation from the regime governed by Stokes' law. The case of $\Gamma = 0.33$ in Fig. 4C displays the strongest amplitude modulation, that is, when a transition to one-dimensional-like swimming occurs (see Fig. 2C) and the swimmer remains in a stable equilibrium state only if $f < f_{so}$. While $\Gamma = 0.50$ also reveals harmonics for 300 kHz, it displays a weak amplitude modulation. It is seen in Fig. 2D that this refers to an unstable equilibrium state for the swimmer that is induced by the confinement independent of the driving frequency $f$.

For driving frequencies sufficiently smaller than the step-out frequency (*i.e.*, $f \leq 0.8 f_{so}$), a helix's angular speed is expected to stabilize to a constant value, that is, when the competing forces reach a balance in an unconfined system (or a sufficiently low degree of confinement). A variation of $\Gamma$ at *f*=300 kHz, see Fig. 5A, shows that this holds qualitatively true for $\Gamma < 0.50$. At $\Gamma = 0.50$, however, the feedback from the boundary walls is strong enough to introduce periodic and non-periodic disturbances to the angular velocity of the helix. Figure 5B plots the swimmer's propulsion speed corresponding to the cases explored in Fig. 5A. It is observed that despite exhibiting qualitatively similar angular speeds for the same driving frequency, there is an increase in $U$ as $\Gamma$ increases.

The Péclet number (*Pe*) is a dimensionless quantity that characterizes the relative importance of advective and diffusive processes. It is defined as the ratio of the rate of advective to diffusive transport. The translational and rotational *Pe* are defined as $Pe_T = vl/D_T$ and $Pe_R = \Omega l^2 / D_R$, respectively. Here, $l$ is the characteristic length scale associated with the object or the flow, $v$ is the characteristic velocity of the fluid flow, and $D_T$ and $D_R$ are the translational and rotational diffusion coefficients, respectively. The Péclet number is thus indicative of whether advection or diffusion dominates in a given system. When $Pe \ll 1$, Brownian diffusion is the dominant transport mechanism, and the system is said to be in the diffusion-controlled regime. On the other hand, when $Pe \gg 1$, advection becomes the dominant transport mechanism, and the system is said to be in the advection-controlled regime. Using LBM simulations, Alcanzare *et al.* demonstrated guided propulsion with nanohelices in the presence of thermal fluctuations that required a $Pe_T > 50$ and $Pe_R > 1$ to achieve an advection-controlled regime with directed motion and thereby effectively overcome Brownian motion (*43*). This is in agreement with the results of Ref. (*55*), wherein based on experimental observations, Ghosh *et al.* concluded that for a helical propeller smaller than a few micrometers in length to achieve unidirectional motion, the driving frequency has to increase as the inverse cube of the swimmer's length.

When the frequency of the driving magnetic field is fixed at 300 kHz, it enables an unbiased comparison of propulsion speeds across the different degrees of confinement. This frequency is sufficiently below the step-out frequency of all systems, and yet high enough to satisfy the $Pe_T > 50$ and $Pe_R > 1$ required for reliably overcoming the effect of thermal fluctuations on the swimmer's trajectories (*43*). It also leads to very small rotational Reynolds numbers $(Re_R \approx 0.001)$ required for creeping flow. Figure 5C illustrates a consistent increase in the rotational drag coefficient $(K_\parallel)$ with an increase in the degree of confinement $(\Gamma)$. The increase in $K_\parallel$ here is observed to be greater for higher viscosities. Figure 5D reports an inverse correlation between the angular speed and $\Gamma$ that is to be expected given the increased drag. However, the average propulsion speed of the swimmer is paradoxically greater in stronger confinement as evident in Fig. 5E. In the next section, an analysis of the fluid pressure as a function of the degree of confinement reveals that this occurs due to an increase in differential pressure when $\Gamma$ increases.

**The principal mechanisms of propulsion in confinement**
Figure 6A summarizes the average differential pressure, $\overline{\Delta q}$, acting on the swimmer in the *y-z* plane as a function of the radial distance from its centerline for different degrees of confinement. As $\Gamma$ increases, $\overline{\Delta q}$ grows pushing the swimmer forward thus making a contribution to its overall propulsion speed. This mechanism overcomes the decrease in coupled-propulsion (*i.e.*, speed of translational motion caused purely by angular speed due to their linear coupling) caused by the increased rotational drag (see Fig. 5C and D), and results in an overall faster forward movement of the swimmer for higher values of $\Gamma$ (see Fig. 5B and E). The magnitude of $\overline{\Delta q}$ close to the swimmer is at its maximum and vanishes exponentially with radial distance away from it. It goes asymptotically to zero at a distance of, approximately, 1.9 times the helix's outer radius across all confinements.

The principal mechanisms that contribute to the swimmer's overall propulsion can be classified into two categories: 1) pure swimming wherein the translational velocity arises solely from the angular velocity of the helical particle as a result of their linear coupling in the low Reynolds regime, and 2) the thrust generated by the confinement from an increased $\overline{\Delta q}$ acting on the particle in the plane perpendicular to its easy axis. Figure 6B shows the contribution of these two mechanisms to the propulsion speed of the swimmer as a function of $\Gamma$ that is quantified by employing the distance traveled per revolution by the swimmer in an unconfined system as a baseline. The contribution of pure swimming shows only a slight decrease when the degree of confinement is increased, whereas the contribution due to differential pressure exhibits a significant rise. For instance, there is about $20\%$ increase in the propulsion speed in the case of

$\Gamma = 0.50$. This effect becomes negligible for $\Gamma < 0.25$ and the coupled translational-angular motion (*i.e.*, pure swimming) is then the only mechanism generating forward motion.

**Modeling the unconfined propulsion speed**
Extensive endeavors, *e.g.*, Refs. (*2, 36–38, 40, 54, 58–66*), have been dedicated to formulating analytical models aimed at elucidating and predicting the hydrodynamics of helical swimmers. Some efforts have resulted in formulations of closed-form equations for different system parameters. For instance, Raz and Avron (*62*), by means of the SBT, concluded that

$$U = \Omega R \frac{\sin(2\alpha)}{3 + \cos(2\alpha)}. \tag{1}$$

This equation implies that the (bulk/unconfined) propulsion speed $U$ acquired by the helix in a creeping flow is independent of viscosity (*i.e.*, in agreement with Fig. 5E and Fig. 6B) and depends linearly on its angular speed $\Omega$ and helical radius $R$, and nonlinearly on the pitch angle $\alpha$ (*i.e.*, the angle between the helix and its centerline). They also concluded that in contrast to a corkscrew moving through a solid medium, a helix requires a minimum of two complete rotations to advance by one lead in a fluid.

De Lima and Moraes (*54*) performed a mechanical analysis of bacterial swimming by treating the helical flagellum as a rigid rotating helix. They decomposed all the forces and velocities acting on the helix into their components normal and parallel to its centerline. By approximating the ratio between the normal and parallel drag coefficients as equal to 2 (*i.e.*, $\xi_\perp/\xi_\parallel = 2$) and multiplying all the contributions (*i.e.*, density forces) by the arc length of the helix (due to acting uniformly on the entire helix), they arrived at an equation similar to Eq. (1). This approximation ($\xi_\perp/\xi_\parallel = 2$) is common in many analytical works in the literature that originate from Ref. (*29*). For instance, there is another relation used, for example, in Ref. (*67*) by taking the same approximation. This relation yielded very similar results as Eq. (1) for all the helices of this work (listed in Table S1). Figure. 7B shows that the propulsion speeds calculated using Eq. (1) are significantly lower than the speeds observed in the simulations. Since these particles exploit drag anisotropy for swimming, the observed differences must be due to the way that these equations account for this factor.

Here, by adopting a similar modeling framework as De Lima and Moraes (*54*), but instead of using $\xi_\perp/\xi_\parallel = 2$ due to the helix possessing cylinder-like uniaxial easy-axis anisotropy, we take the ratio as equal to the ratio of the corresponding projected surface areas of a cylinder with the same aspect ratio and diameter as the helix. Thus, we arrive at a new equation to calculate the propulsion speed of the swimmer,

$$U = \Omega R_o \frac{(A^{-1} - 1)\sin(2\alpha)}{A^{-1}(\cos(2\alpha) - 1) - \cos(2\alpha) - 1}. \tag{2}$$

Here, $A = 2nP/\pi R_o$ in which $n$ and $P$ are, respectively, the number of helical turns and pitch length. A consequence of Eq. (2) is that given $\Omega$ and $R_o$, the pitch angle which maximizes propulsion speed depends on $A$ and is given by $\alpha = \arctan(\sqrt{A})$, which favors a more upright pitch angle when the number of turns increases.

Using Eq. (2) to calculate the ratio of the swimming speed of a helix to its linear speed when penetrating a solid, gives $(1 - A^{-1})/2\pi(\cot^2(\alpha) + A^{-1})$. This implies that in the Stokesian regime, for example, swimmer H0 needs at least 12.5 rotations to advance the distance of one pitch in fluid and that this is independent of viscosity. This aligns well with the results of previous computational studies of bacterial swimming using BEM indicating that more than 10 rotations of the flagellum are needed to propel the organism forward a distance of one helical wavelength (*33*).

The predictions of both Eqs. (1) and (2) are reported by the solid lines in Fig. 7B for swimmer H0 and Fig. S2 for swimmers H1-H8 (Table S1 lists the swimmers' characteristics). These results show that the propulsion speeds calculated by our new equation, *i.e.*, Eq. (2), are significantly closer to the simulation results for the unconfined systems $(\Gamma = 0)$ when compared against the results of Eq. (1). Equation (2) yields outcomes which are the closest to the simulation results for particle H0 followed by H1. This has to do with the particles' $R_m/R_M$ (where $R_m < R_M$), which is 0.5 for H0, 0.2 for H1, and < 0.2 for the rest. In the derivation of Eq. (2), in order to determine $\xi_\perp/\xi_\parallel$, we approximated the helix by a cylinder and, therefore, particles with a greater value of $R_m/R_M$ better fit that assumption. It is important to note, however, that the applicability of both Eqs. (1) and (2) is restricted to unconfined swimmers.

**Predicting the confined propulsion speed**
Analysis of the pairwise relationships among $\Omega, U$, and $f$ (*i.e.*, the simulation data points in Fig. 7) for $f \leq 0.8 f_{so}$ reveals their direct linear correlations in agreement with Eq. (2) as well as experiments, e.g., Ref. (*39*), that are consistent across all values of $\Gamma$. However, as the level of confinement increases, $U$ exhibits a sharper growth as a function of both $\Omega$ and $f$ (see Fig. 7B and C). An exponential regression of the gradients of linear interpolation lines connecting the simulation data points in Fig. 7B as a function of $\Gamma$ shows that the confined propulsion speed of the swimmer $(U_c)$ can be written as a function of the propulsion speed in the absence of confinement $(U_\infty)$ and the ratio between the characteristic hydrodynamic length scales of the system using the following relation,

$$U_c = U_\infty e^{\Gamma^b} = U_\infty e^{\left(2\frac{R_M+R_m}{L_z}\right)^b} \qquad (3)$$

With *b* = 2.5. This relation recovers the presence of nonlinearity when confinement is involved regardless of the driving frequency (see the dotted line in Fig. 5E). Figure 7B illustrates the confined propulsion speed of the swimmer calculated by Eq. (3) for $\Gamma$ as a function of $\Omega$ that are in great agreement with the simulation results. Figure. 7 reports results for particle H0 (see Table S1). To evaluate the generality of Eq. (3), swimmers H1-8 (see Table S1) were created by varying different parameters, *i.e.*, major and minor radii, pitch angle and density, and were examined using two different degrees of confinement, *i.e.*, $\Gamma = 0.29, 0.40$. The dashed and dotted lines in Fig. S2 present the predictions of Eq. (3) for these helices showing excellent agreement with the simulations. Using the simulation data of all swimmers (*i.e.*, H0-8) for fitting to estimate $b$ gives $b = 2.5 \pm 0.1$. It is evident in Figs. 7B and S2 that Eq. (3) with $b = 2.5$ fits every simulated scenario very well.

In Poiseuille flows, the pressure decreases along the *x*-direction to balance the viscous force (*68*). For the case of a channel with a rectangular cross-section and the *y*-dimension being much larger than the *z*-direction (*i.e.*, the conditions considered in this work), relative differential pressure scales (ideally) by $\Gamma^2$ for the non-driven system. However, in a driven system, the fluid continuously receives additional energy per unit volume from the kinematics of the particle. It is thus plausible that the exponential in Eq. (3) accounts for this effect.

The robustness of Eq. (3) in effectively predicting propulsion speed for the impact of confinement offers significant advantages. The results consistently demonstrate its reliable performance across various scenarios and configurations. By accurately modeling the interaction between propulsion and confinement, this heuristic tool can be foreseen to be instrumental in developing applications such as targeted drug delivery in tissues, veins, and lab-on-a-chip nano/microfluidic devices.

For instance, replacing the computationally estimated $U_\infty$ in Eq. (3) with the improved analytical expression for the speed of an unconfined helix, *i.e.*, Eq. (2), yields

$$U_c = \Omega R_o \frac{(A^{-1}-1)\sin(2\alpha)}{A^{-1}(\cos(2\alpha)-1)-\cos(2\alpha)-1} e^{\left(2\frac{R_M+R_m}{L_z}\right)^b}, \tag{4}$$

which makes it possible to estimate (see Fig. S3) the host environment-specific swimming speed of a helix by considering the degree of confinement without requiring any computational or experimental tests. This expression can be easily employed for engineering systems involving the application of these swimmers in constricted fluidic environments such as lab-on-a-chip devices.

**DISCUSSION**

Life at low Reynolds numbers commonly refers to Stokesian dynamics based on a linearized version of the Navier-Stokes equations that neglects advective inertial forces, assuming that the system does not retain any memory of its past (*1–3*). Simulation methods such as RFT, SBT and BIF based on these assumptions have enjoyed a measure of success in modeling the swimming dynamics of biological or bioinspired microswimmers in a regime wherein the driving frequencies are invariably less than one kHz, *e.g.*, Refs. (*28, 34, 35, 38, 53, 55, 67, 69–74*). However, controlled steering, propulsion and separation of magnetic helical nanoswimmers require operation in a frequency regime that is almost 2–5 orders of magnitude greater, from a few hundreds of kHz to MHz (*42, 43, 55*). In this regime, our results obtained using LBM-based full-scale nonlinear hydrodynamic simulations of magnetic helical nanoswimmers show that the aforementioned defining assumptions of Stokesian dynamics do not always hold, especially under confinement wherein our results show nonlinear behavior.

Briefly, we have investigated the dynamics of nanoscale magnetic helical filaments driven by an external rotating magnetic field in a fluid under confinement using LBM to study the underlying physical mechanisms of their propulsion. The motion of helical filaments exhibited local nonlinear dynamics even for driving frequencies well below the step-out frequency $f_{so}$ of the system. Furthermore, independent of the degree of confinement, dominant nonlinear behavior was observed for frequencies close to and above the step-out frequency wherein the propulsion speed could no longer be described as a linear function of the angular speed. Steady and linear dynamics were recovered over longer time scales on average for $f < f_{so}$ consistent with the assumptions and predictions of the Stokes equations. This also explained the basis of the partial limited success enjoyed so far by the simulation methods such as RFT, SBT and BEM that assume Stokesian dynamics in explaining often contrasting experimental and computational results (*21*, *28*). However, nonlinearities are present in the dynamical behavior of nanohelices independent of the driving frequency $f$ at higher degrees of confinement, and they are accompanied by an emergence of a transition to one-dimensional swimming at a critical degree of confinement $(\bar{\Gamma} = 0.33)$.

Here, we have proposed an improved equation, *i.e.*, Eq. (2), to model the coupling between the translational and angular speeds of an unconfined helix. The swimming speeds calculated using this closed-form relation yielded significantly closer results to simulation when compared against the propulsion speeds obtained from similar previous relations (*54*, *62*, *67*). We then proposed a new empirical relation, *i.e.*, Eq. (3), to model the nonlinear dependence of the propulsion speed of a helical filament in confinement. The empirical relation in Eq. (3), with an exponential term is based purely on the geometric degree of confinement. It accurately predicted the nonlinear propulsion speed for an extensive set of geometric variations in the shapes of the nanoswimmers, regardless of the driving frequency of the rotating external magnetic field. As such, the proposed empirical relation provides an important practical tool for the design of systems seeking to exploit the use of driven helical swimmers in constricted fluidic environments. Additionally, replacing the computationally estimated $U_\infty$ in Eq. (3) with the analytical expression for the speed of an unconfined helix, *i.e.*, Eq. (2), results in a purely analytical expression, *i.e.*, Eq. (4), for the propulsion speed of a helical nanoswimmer. This analytical expression yields a significantly improved estimate for the effect of confinement that can be easily employed for engineering designs of optimal systems to enhance performance and driving advancements in the field of small-scale swimmers without resorting to any expensive computations or experimentations.

An increase in the rotational drag was observed with an increase in the degree of confinement $\Gamma$. Paradoxically, however, the swimmers achieved (exponentially) faster propulsion in tighter confinements because of the differential pressure created by the boundary. The contribution from the differential pressure to the swimmer's overall propulsion speed increased with $\Gamma$. However, this effect due to confinement diminished as the channel width was increased and nearly vanished when the channel width exceeded four times the hydrodynamic diameter of the particle. Thus, we have identified a clear lower bound on the degree of confinement beyond which the propulsion of nanohelices is primarily characterized by pure swimming and the hydrodynamic feedback from the bounding walls can be neglected. In this regime of pure swimming, the propulsive motion of a helical nanoswimmer results solely from the coupling between the translational and rotational velocities of the corkscrew-like particle in a Stokes flow.

**METHODS**

Small-scale swimmers are broadly classified as catalytically (*75–77*) or electromagnetically (*74*, *78–81*) driven depending on the source of their motive force. The catalytically driven swimmers exploit chemical gradients for propulsion and require a fuel such as hydrogen peroxide, acidic or alkaline solution. However, electromagnetically driven swimmers are better candidates for *in vivo* applications as they are not sensitive to chemical gradients and do not require a fuel. In addition, most biological tissues are typically regarded as diamagnetic. For instance, human tissue is weakly diamagnetic with a small volume (static) magnetic susceptibility of the order of $10^{-6}$ (SI units), which allows the magnetic field to penetrate into it (*82*, *83*). However, magnetic susceptibility of a tissue can vary depending on the frequency range of the applied magnetic field (*84*). At frequencies in the kHz range, magnetic susceptibility is generally considered to be independent of frequency (*85–87*). This is often referred to as the static (or low-frequency) magnetic susceptibility. In this regime, the magnetic properties of tissues are primarily determined by the presence of naturally occurring paramagnetic and diamagnetic substances, such as oxygen, iron, and water (*88*). At higher frequencies, especially in the radio frequency (RF) and microwave ranges (MHz to GHz), additional factors such as electrical properties of the tissue and the relaxation processes of water molecules and ions come into play. As a result, tissues exhibit a combination of diamagnetic and paramagnetic properties that can be influenced by frequency (*85–87*). For example, in magnetic resonance imaging (MRI), the frequency dependence of the magnetic susceptibility is exploited to generate contrast and obtain detailed images of different tissues in the body. By applying specific RF pulses at different frequencies and measuring the resulting signal response, MRI can differentiate between different types of tissues and provide detailed structural and functional information (*82*).

Additionally, the relative electric permittivity of a typical biological tissue also depends (strongly) on the frequency and can be orders of magnitude larger than that of water at frequencies in the kHz range (*i.e.*, β-dispersion), eventually leveling to the order of $10^2$ beyond the 10 MHz range (*85, 89, 90*). In this work, the field frequency for effective electromagnetic propulsion of the under-study particles (see Table S1) in water is 300 kHz. At this frequency, the magnetic susceptibility remains in the static regime and the high dielectric permittivity of the human tissue effectively screens out the external electric field. As such, the field frequency considered in this study is suitable for electromagnetic steering of helical nanoparticles in living tissue.

LBM (*48*) was employed in the single relaxation time Bhatnagar-Gross-Krook approximation (*91*) to simulate fluid flow using a conservative coupling (*92*) to the helical nanoparticles. LBM solves the discrete Boltzmann transport equation on a structured lattice to emulate incompressible fluid. The Knudsen (*Kn*) and Mach (*Ma*) numbers are two important dimensionless parameters in LBM. The *Kn* describes the ratio of the molecular mean free path (*i.e.*, the average distance a molecule travels between collisions with other molecules) to the characteristic length scale. It is employed to assess the appropriateness of using a continuum or rarefied gas model to simulate fluid flow as it provides a measure for the significance of rarefied gas effects that occur when the mean free path becomes comparable to or larger than the characteristic length scale of the flow system. For $Kn \ll 1$, the gas behaves like a continuum, and continuum fluid dynamics equations, such as Navier-Stokes, can be applied. For flow simulations in the continuum regime the LB equation is solved using a lattice with a sufficiently small lattice spacing. This approach relies on the assumption that molecular collisions dominate, and the fluid can be described using macroscopic fluid variables such as density and velocity.

*Ma* is the ratio of the flow velocity of a fluid to the speed of sound in the fluid. It is used to determine whether the flow is compressible or incompressible. In LBM, the incompressible flow regime is commonly approximated by setting a limit on *Ma*. For most practical purposes, a *Ma* below 0.3 is considered a good approximation for an incompressible flow (*93*). This implies that the flow velocity is significantly lower than the speed of sound, resulting in minimal density variations within the fluid. As a result, the density can be assumed constant and LBM recovers solutions of the Navier-Stokes equations in the limit of low *Kn* and *Ma* (*46*).

Another important dimensionless parameter is the Reynolds number (*Re*), which characterizes the flow regime of a fluid, particularly in relation to the relative importance of inertial to viscous forces. The translational and rotational *Re* are calculated using $Re_T = \rho v l/\mu_d$ and $Re_R = \rho \Omega l^2/\mu_d$, respectively. $\rho$ is the density of the fluid. The *Re* indicates the type of flow regime, *e.g.*, laminar, transitional, or turbulent. When $Re \ll 1$, the flow is smooth and orderly wherein viscous forces dominate.

This work employs a hybrid Lattice Boltzmann-Molecular Dynamics (LB-MD) scheme (*42, 48, 92, 94*), that incorporates all viscous and inertial effects, to investigate the effects of the degree of confinement on the propulsion of rigid helical nanofilaments (Fig. 8A) propelled by a rotating magnetic field. In the systems studied, *Ma* ranges from $5 \times 10^{-6}$ to $2 \times 10^{-5}$ ($v_s \approx \sqrt{0.2}\Delta x/\Delta t$, where $v_s$ is the speed of sound in the system and $\Delta x/\Delta t$ is the lattice velocity (*48*)). A field frequency of $300 \text{ kHz}$ is chosen for a comparison of propulsion across different values of $\Gamma$. This ensures that, in all cases, the frequency stays well below the step-out frequency and that the fluid regimes remain largely dominated by viscous forces. This leads to $Re_R \approx 0.001$ with slight differences accruing from the different degrees of confinement. Figure 8B shows the rotational Reynolds number as a function of field frequency for different $\Gamma$ and $\mu_d$.

All simulations were performed in LAMMPS (*95*) using the D3Q15 LB model described in Refs. (*48, 92, 94*) to accurately account for the interparticle hydrodynamic interactions and the particle-fluid interfacial coupling using a no-slip boundary condition (*96*). The notation D3Q15 indicates a 3D lattice with 15 velocities. In order to couple a moving object to the LB fluid, a discretized representation of the object on the fluid lattice is required. This is accomplished by dividing the surface of the object into a set of nodes, and then distributing the coupling forces from these nodes to the nearby lattice sites. This was done using the Peskin stencil (*94*), which is based on the immersed boundary method (*97, 98*). It uses a smoothing kernel (*i.e.*, a 4-point approximation to the Dirac delta function) to spread the influence of a point particle over a limited region (*i.e.*, compact support of 64 grid points). This coupling consistently introduces forces on both the nodes and the fluid resulting from an energy and momentum conserving interaction.

The helical nanoswimmers (listed in Table S1) were modeled as rigid shells with uniformly distributed surface (pseudo-)atoms that interact with the LB fluid lattice sites. Except for H8 (see Table S1), the surface atom masses were selected to ensure equivalence of the total mass of the swimmers and the mass of the displaced fluid such that they are neutrally buoyant. Additionally, the number of surface atoms for the swimmer also constrains the surface area per atom for the shell to be smaller than the square of the lattice spacing ($\Delta x^2 = 16 \text{ nm}^2$) to avoid spurious errors from an insufficiently discretized mesh (*99*). Periodic boundary conditions were employed in the *x*- and *y*-directions, whereas the LB bounce-back boundary along with purely repulsive top and bottom Lennard-Jones (LJ) potential walls were specified in the *x-y* plane to obtain a no-slip boundary condition for the fluid flow and to emulate two-dimensional (2D) confinement of the swimmer, respectively. The simulation box dimensions in the *x*- and *y*-directions were taken to be 8 times the axial length ($L_a$) and outer diameter ($D_o$) of the swimmer, respectively; tests using dimensions 4 times larger showed no change in the results. $D_0$ is related to the major and minor radii via $D_0 = 2 \times (R_M + R_m)$, see Table S1. To simulate different degrees of confinement ($\Gamma$), the distance between the *z*-walls ($L_z$) was varied as a multiple of ($D_0$). In the case of an unconfined swimmer, the size of the simulation box in the $z$-direction was set to $10D_0$. Fluid viscosities of one fourth, half, and equal to that of water at room temperature corresponding to simulation time steps of 21.30, 10.65, and 5.32 ps, respectively, were considered to quantify and understand the effect of momentum diffusion in the fluid and its impact on the propulsion velocity of the externally driven propeller under confinement.

In the case of a rotating rigid body in a fluid, the drag force exerted by the fluid on it can be resolved into translational and rotational components. Then, the linearity of the Stokes equations at very low Reynolds numbers allows representation of these forces in terms of resistance matrices acting on the linear velocity $\mathbf{U}$ and angular velocity $\mathbf{\Omega}$ of the rigid motion (69),

$$\begin{bmatrix} \mathbf{F} \\ \mathbf{T} \end{bmatrix} = \begin{bmatrix} A & B \\ B & C \end{bmatrix} \begin{bmatrix} \mathbf{U} \\ \mathbf{\Omega} \end{bmatrix}. \qquad (5)$$

Here, $\mathbf{F}$ and $\mathbf{T}$ are the applied force and torque required to pull the helix with linear velocity $\mathbf{U}$ and angular velocity $\mathbf{\Omega}$. The terms $A$, $B$, and $C$ are $3 \times 3$ resistance matrices that only depend (nonlinearly) on the geometry of the swimmer. Unlike bodies with spherical symmetry that have zero off-diagonal entries in their resistance matrix, Eq. (5) implies that for corkscrew-shaped swimmers the translational and rotational motions are coupled due to the spontaneous viscous drag symmetry breaking by the helical particle (50). In other words, rotational velocity causes translational velocity and vice versa (3, 69). This entertains the idea of using a continuously rotating magnetic field to propel the chiral particle around its long axis in fluid to gain linear velocity, and therefore, controlled locomotion.

It is important to note that chirality is not essential for propulsion. Symmetries (*e.g.*, chirality) of a driven object cannot be determined solely by its geometry and symmetry properties of the magnetic dipolar moment affixed to it must be considered (100). For instance, if instead of a chiral structure with a permanent dipole, a polarizable achiral planar object with an induced electric dipole is considered, it can gain unidirectional motion by electrorotation quite efficiently owing to a spontaneous symmetry breaking, whereas the structure remains yet achiral (101, 102).

Assuming a steady-state and uniform rotating magnetic field, the external magnetic force $\mathbf{F}$ in Eq. (5) vanishes and the magnetic interaction can be modeled by considering only a magnetic torque $\mathbf{T}$ generated by the interaction of the swimmer's permanent magnetic dipole moment $\mathbf{M}$ with the external magnetic field with a magnetic flux intensity of $\mathbf{B}$ (see Fig. 8A) (53). The magnetic torque is calculated using $\mathbf{T} = \mathbf{M} \times \mathbf{B}$, where $\mathbf{B} = \sin(ft)\hat{\mathbf{y}} + \cos(ft)\hat{\mathbf{z}}$, $f$ is the field frequency, and $t$ is the simulation time step (42). Hence, the swimmer experiences maximal $\mathbf{T}$ when its long axis is perpendicular to the imaginary plane in which the magnetic field vector rotates. The experimental value of the magnetic dipole moment of the particle is estimated to be $2 \times 10^{-17}$ Am$^2$ (103). For $MB = 1.5 \times 10^{-18}$ Nm in the simulations (43), the magnetic field strengths required for experimenting will be 75 mT.

**Supplementary Materials**
   Figs. S1 to S3
   Table S1

**Acknowledgments:** The authors thank Dr. Colin Denniston from the Department of Physics and Astronomy at Western University for his support with the simulation technique used. M.K. acknowledges financial support from the Natural Sciences and Engineering Research Council of Canada (NSERC) and Canada Research Chairs Program. Computational resources were provided by the Digital Research Alliance of Canada.

**Funding:**
Natural Sciences and Engineering Research Council of Canada Discovery Program (MK)
Canada Research Chairs Program (MK)

**Competing interests:** Authors declare that they have no competing interests.

**Figures:**

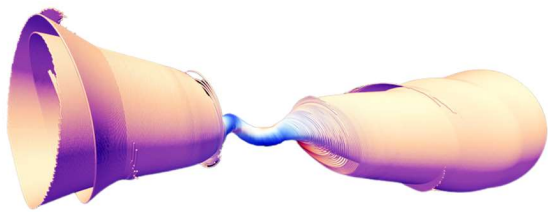

**Graphical abstract.** Perspective view of a left-handed helical filament (in middle) swimming in water along with the conical fluid streamlines created at the downstream (left) and upstream (right) ends.

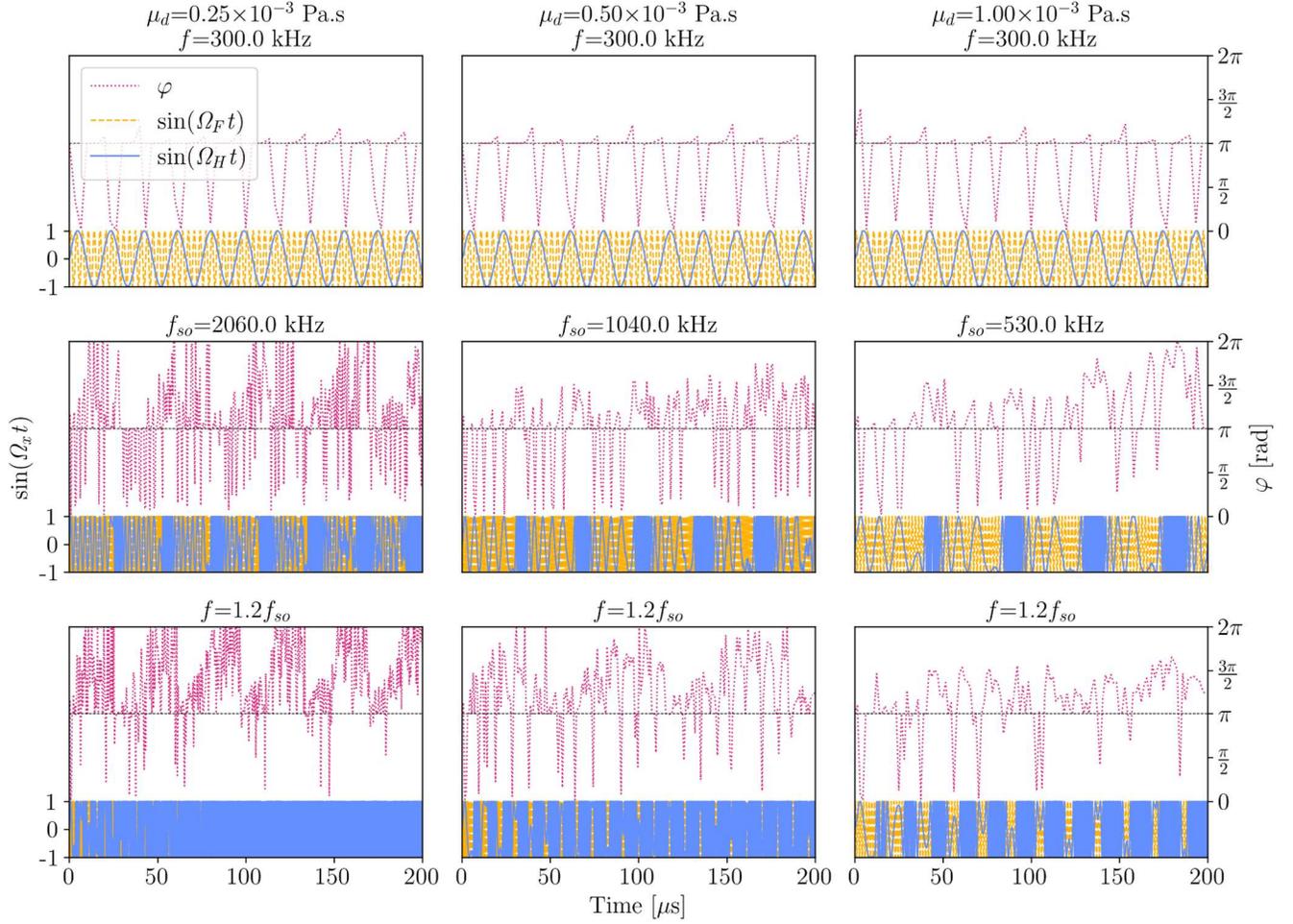

**Fig. 1. Phase slips between the rotation of the magnetic field and the resulting rotation of the swimmer.** Orange dashed, solid blue, and pink dotted lines denote, respectively, the sine of field's angular speed, sine of swimmer's angular speed, and phase difference between the two in radians $(\varphi)$ at $\Gamma = 0.2$. The first, second, and third rows correspond to driving frequencies, respectively, below, at, and higher than $f_{so}$. The left, middle, and right columns correspond to fluid viscosities $(\mu_d)$ of one fourth, half, and equal to that of water, respectively. When the driving frequency is smaller than $f_{so}$, the swimmer remains synchronized with the field (top row). At $f_{so}$ and beyond (middle and bottom rows), the magnetic torque is not strong enough to overcome the increased viscous drag torque to keep the rotation of the swimmer in harmony with the field; thus, episodes of phase slip emerge and the swimmer falls behind the magnetic field. This decreases the average angular speed of the swimmer significantly as reported in Fig. S1.

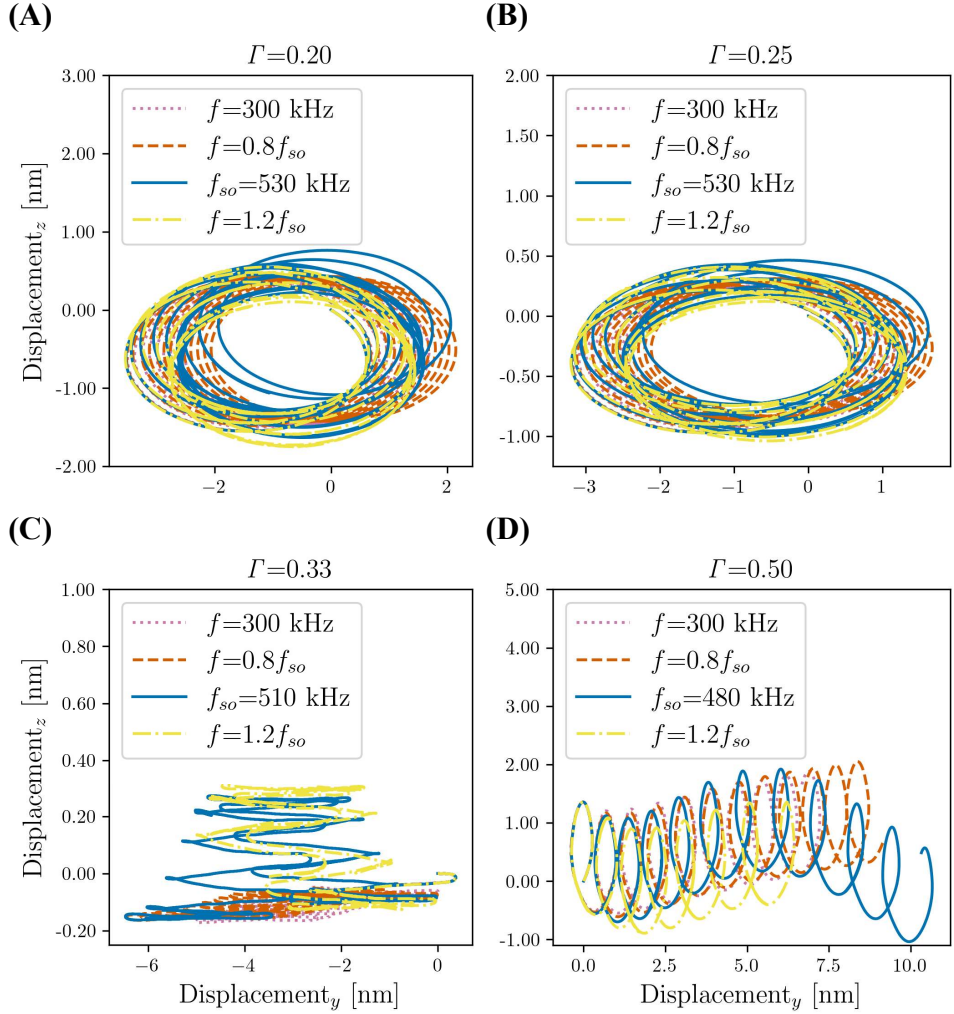

**Fig. 2. Swimmer's trajectories in the plane perpendicular to the axis of travel.** Swimmer H0's trajectories (see Table S1 for its geometrical properties; the distance between the walls depends on $\Gamma$ and $D_0$ via $L_z = D_0/\Gamma$) in the $y$-$z$ plane under different degrees of confinement $\Gamma$ when driven by field frequencies below, at, and above the step-out frequency at a fluid viscosity equal to that of water. For $\Gamma > 0.25$, the feedback from the boundary walls is strong enough (see also Fig. 3) to effectively disrupt the swimmer's $y$-$z$ trajectory, resulting in a deviation from the trajectory anticipated by theory. For the case of $\Gamma = 0.33$ [*i.e.*, **(C)**], when the particle is driven by field frequencies below the step-out frequency, however, the range of changes in the $y$- and $z$-displacements are reduced significantly and a transition to one-dimensional-like swimming emerges. Under a stronger confinement, *e.g.*, $\Gamma = 0.50$, the feedback from the side walls shifts the swimming directionality, pushing the swimmer sideways in a direction opposite to its chirality during its forward motion.

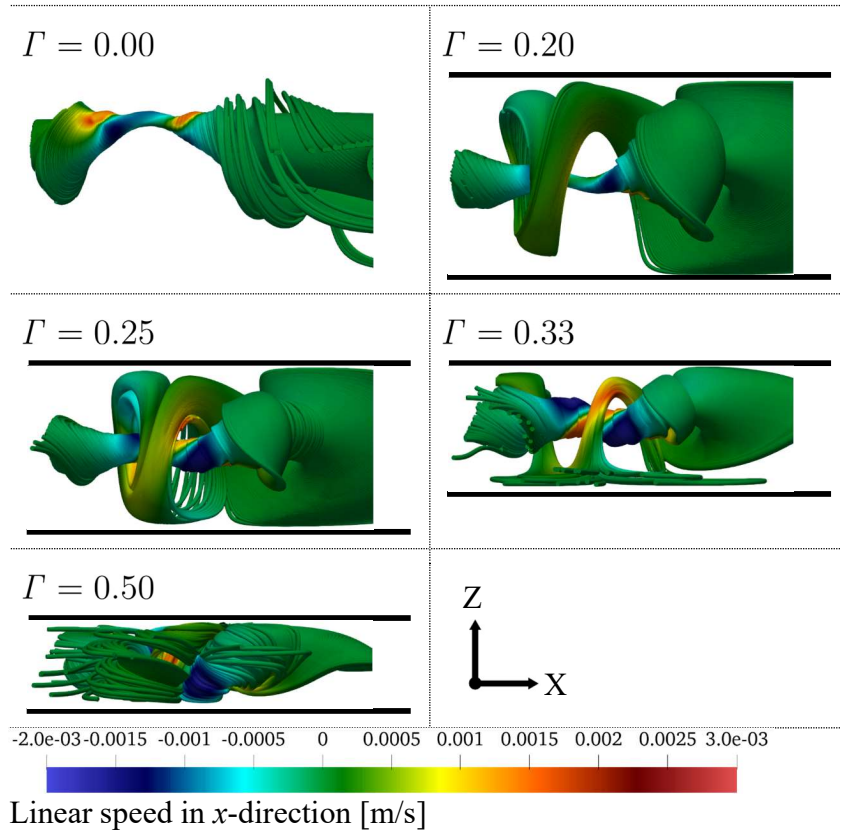

**Fig. 3. Fluid streamlines created around the swimmer under different degrees of confinement.** Fluid streamlines created around swimmer H0 (see Table S1) for $\Gamma$ at $t = 15\,\mu s$, which is enough for the systems to reach steady state. In the absence of confinement (*i.e.*, $\Gamma = 0$) fluid streamlines are created only at the downstream (left) and upstream (right) ends of the swimmer. Confinement leads to fluid streamlines at the ends and around the swimmer's centerline. As $\Gamma$ increases, the proximity of the swimmer to the boundary walls generates greater feedback, giving rise to a stronger boundary effect, *e.g.*, at $\Gamma = 0.50$. The tiles are cropped along the *x*-axis for clarity; their aspect ratios do not represent the true aspect ratios of the simulation box. Visualized using ParaView (*56*).

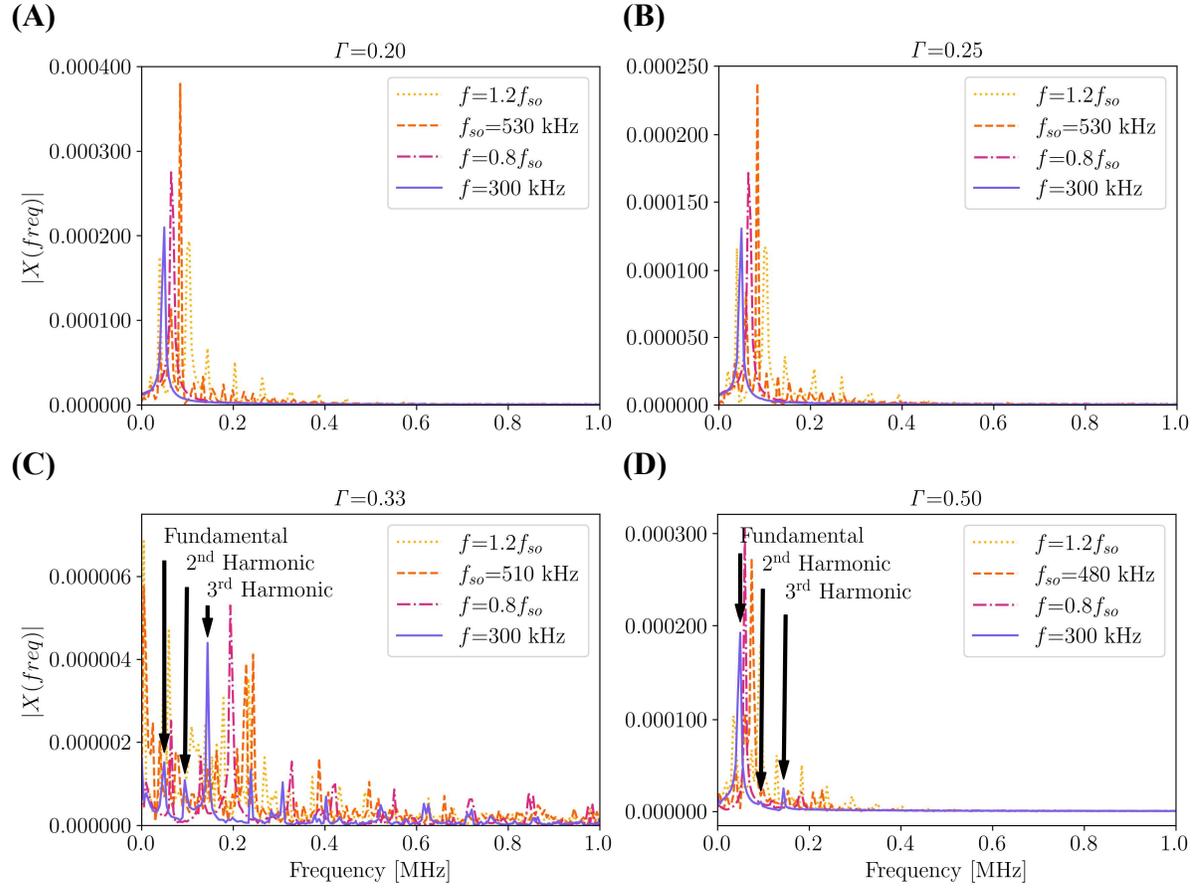

**Fig. 4. Discrete Fourier transform of linear velocity in the direction perpendicular to the boundary walls.** Fast Fourier Transform (FFT) analyses of the translational velocity for the swimmer H0 (see Table S1) in the direction perpendicular to the confinement walls are reported for different values of $\Gamma$. For each degree of confinement, the particle is driven at field frequencies below, at, and above the step-out frequency as illustrated by the different line styles. For $\Gamma > 0.25$, the spectrums reveal harmonics for all driving frequencies examined including $300\,\text{kHz}$, which is the base driving frequency of this work and significantly below $f_{so}$. The spectrums do not show harmonics for the base frequency when $\Gamma \leq 0.25$.

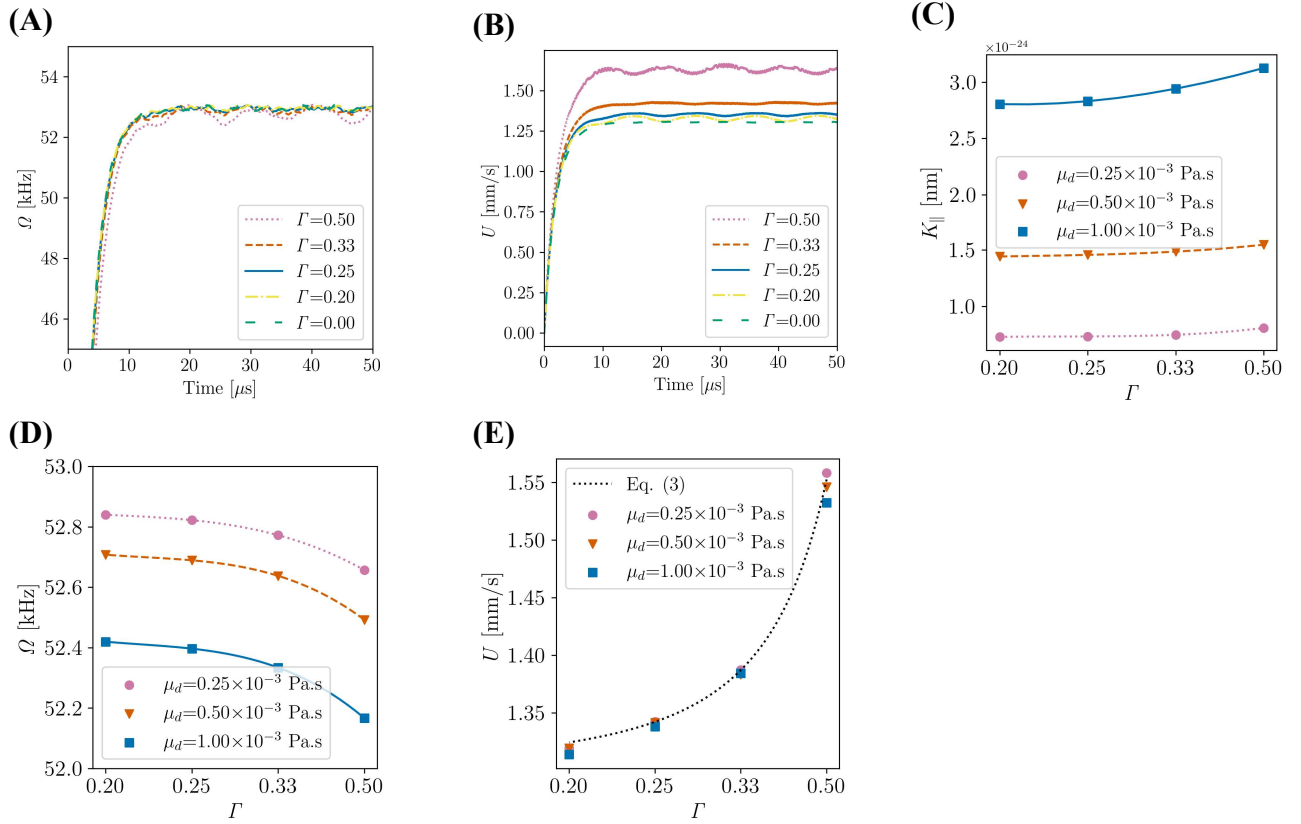

**Fig. 5. Swimmer's speeds and rotational drag coefficient.** (A) angular speed, (B) propulsion speed, (C) rotational drag coefficient, (D) average angular speed, and (E) average propulsion speed of swimmer H0 (see Table S1) for different degrees of confinement driven at $300\,\text{kHz}$. In narrow channels with higher $\Gamma$, the swimmer faces a greater rotational drag, decreasing its angular speed, but at the same time, it exhibits a faster propulsion speed on average.

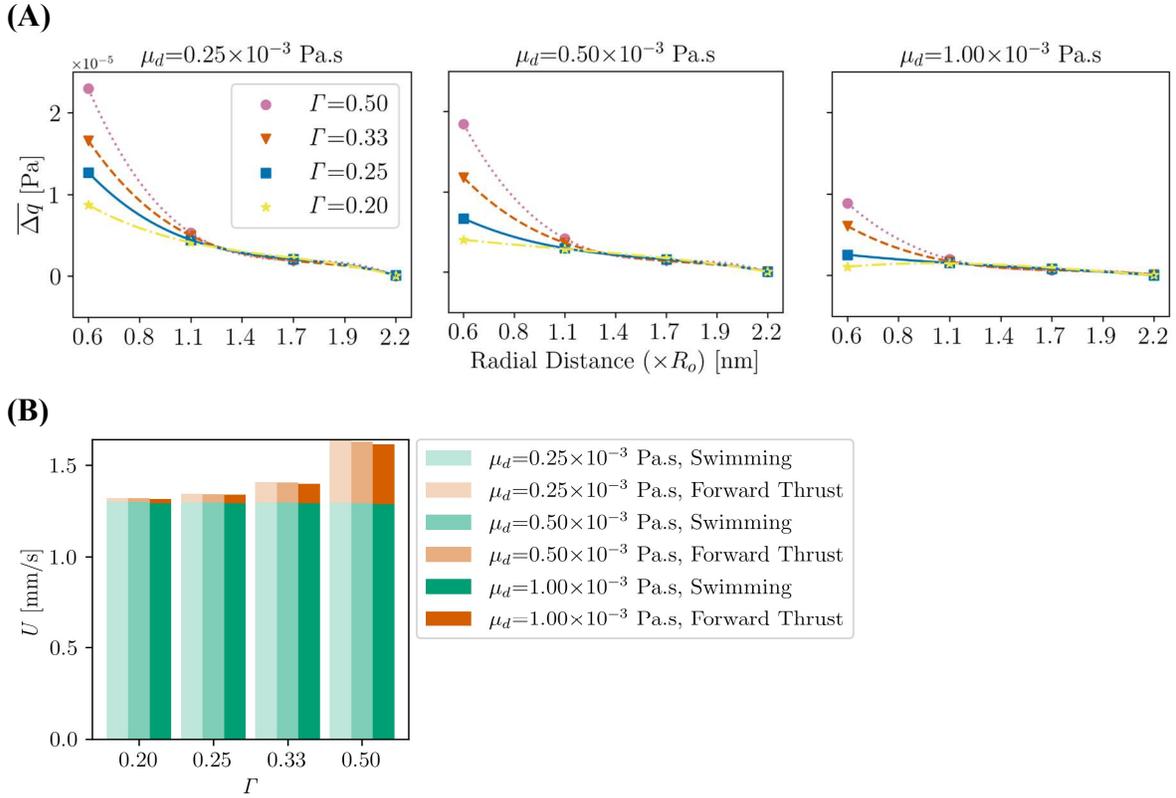

**Fig. 6. Differential pressure and total propulsion. (A)** Average net dynamic pressure, $\overline{\Delta q}$, acting on the swimmer H0 in the *y-z* plane as a function of radial distance normalized by the helix's outer radius (*i.e.*, $R_o = R_M + R_m$, where $R_M$ and $R_m$ are, respectively, the helix's major and minor radii, see Table S1) from its centerline for different degrees of confinement and fluid viscosities of one fourth, half, and equal to that of water. As the degree of confinement is increased, the resulting net pressure amplifies on average, leading to a more substantial contribution to forward thrust. **(B)** The overall propulsion speed of H0 is decomposed into contributions from the two principal mechanisms causing forward motion (*i.e.*, pure swimming and the push created by the confinement) as a function of $\Gamma$ for fluid viscosities of one fourth, half, and equal to that of water. As $\Gamma$ is increased, the share of $\overline{\Delta q}$ grows resulting in a net faster propulsion of the swimmer. The viscosity-independence of swimming speed in the Stokesian regime is evident here.

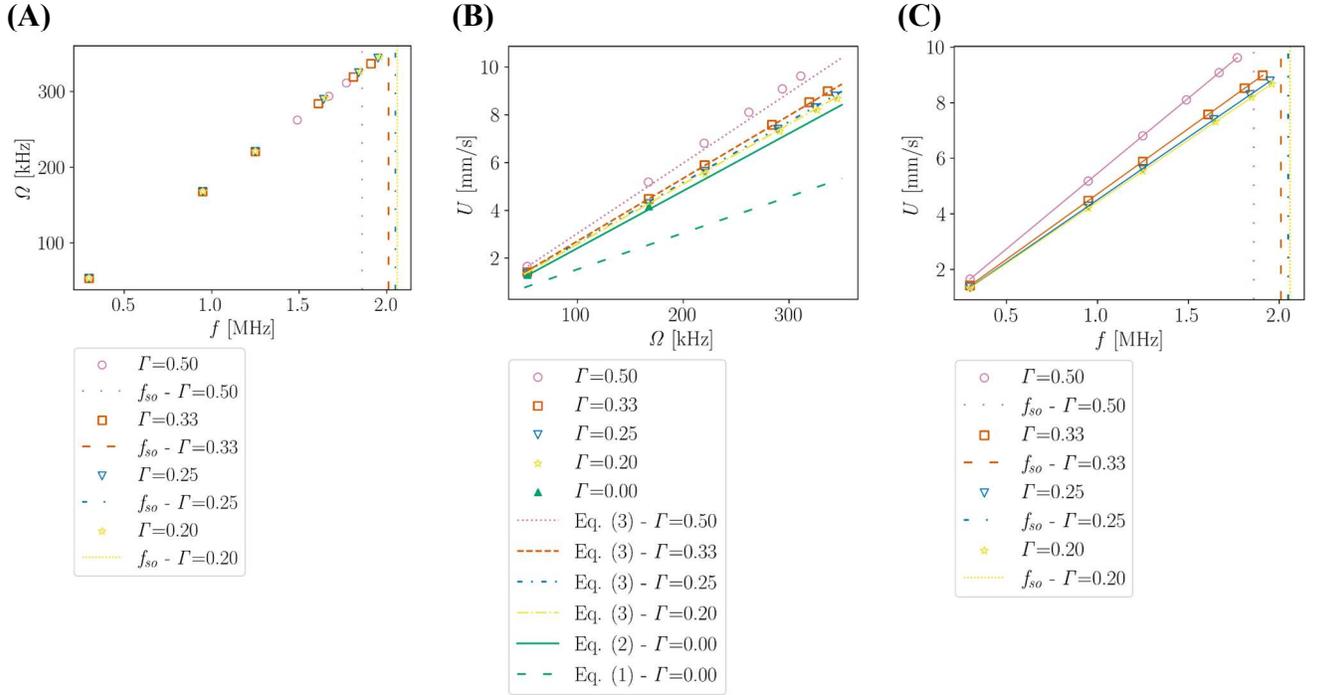

**Fig. 7. Pairwise relationships among angular speed, propulsion speed, and driving frequency.** $\Omega$, $f$, and $U$ display strong linear pairwise correlations for $f \leq f_{so}$. The data points represent simulation results for different values of $\Gamma$ for H0. The line plots in **(A)**, **(B)**, and **(C)** are, respectively, the corresponding $f_{so}$ for each $\Gamma$, predictions of the equations, and interpolation. **(B)** illustrates that Eq. (3) with $b = 2.5$ accurately captures the effect of increasing confinement on the propulsion of the swimmer. Also, the propulsion speeds in the absence of confinement calculated by Eq. (2) are significantly closer to the simulation results for $\Gamma = 0$ than those of Eq. (1).

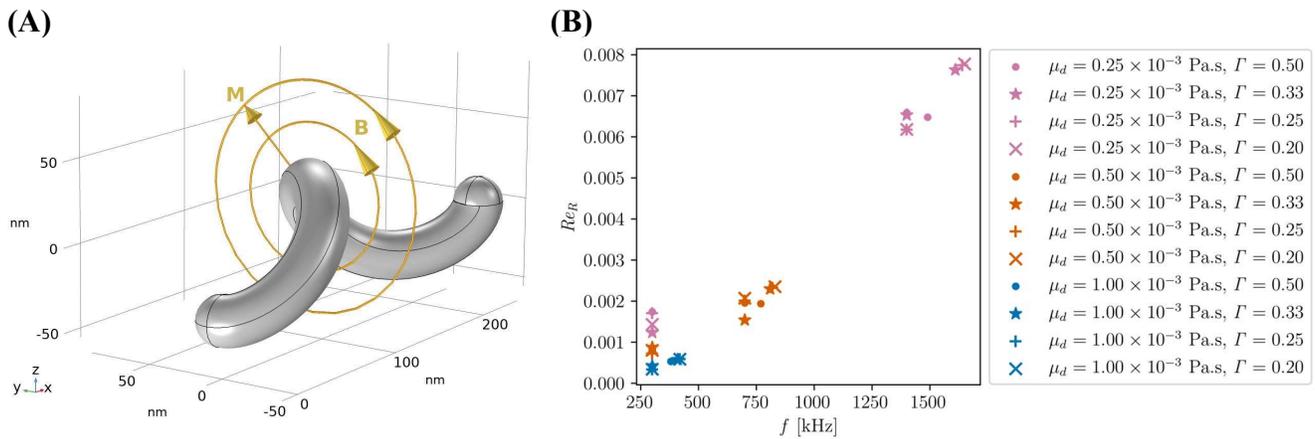

**Fig. 8.** **(A) Vector representation of the magnetic interaction.** The permanent magnetic dipole moment **M** of the particle is fixed perpendicular to its long axis. The interaction of **M** with the rotating external magnetic field with a magnetic flux intensity of **B** generates a torque vector **T** around the particle's centerline. Table S1 lists the properties of all helical filaments used in this work. **(B) Reynolds number as a function of field frequency.** For $f < f_{so}$, as $f$ increases, the angular speed of the swimmer also increases, yielding a greater $Re_R$. Pink, orange, and blue markers pertain to cases with $\mu_d$ of one fourth, half, and equal to that of water, respectively.

# Supplementary materials

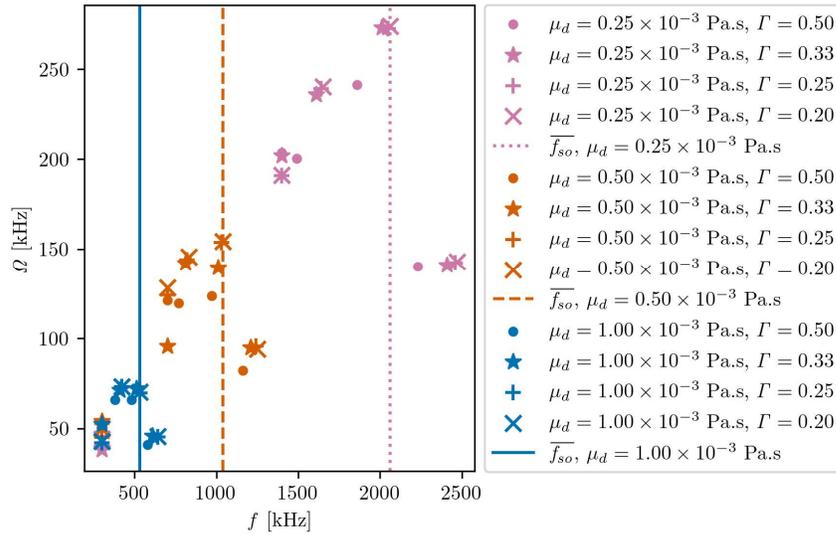

**Fig. S1. Driving frequency versus angular speed.** The angular speed of swimmer H0 (see Table S1) is depicted as a function of the frequency of the external magnetic field. Pink, orange, and blue colors correspond to fluid viscosities of one fourth, half, and equal to that of water, respectively. The vertical lines indicate the average step-out frequency $(\overline{f_{so}})$ across the different degrees of confinement relevant to each viscosity. Circle, star, plus, and cross markers denote the average angular speed of the swimmer for different degrees of confinement $\Gamma$. For driving frequencies well below the step-out frequency $(< 0.8\overline{f_{so}})$, the average angular speed increases linearly with frequency. As approaching or once past the step-out frequency, the average angular speed drops sharply due to the phase slips/flips reported in Fig. 1, consequently, the swimmer becomes increasingly inefficient in generating forward motion.

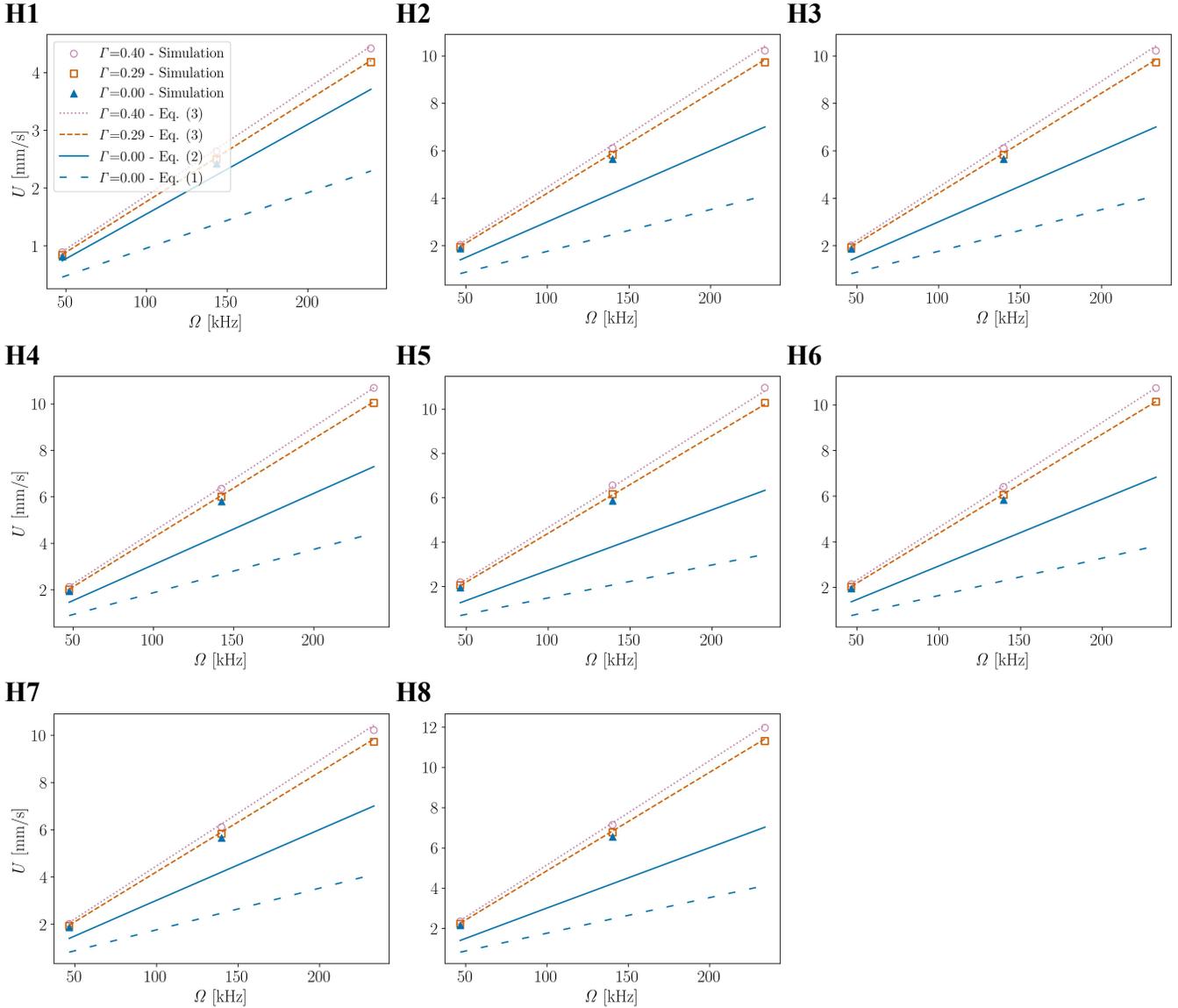

**Fig. S2. Relationship between propulsion and angular speeds for various helical nanoswimmers Hx.** To examine the applicability of Eq. (3) in accurately predicting propulsion for the effect of confinement, all the remaining particles (*i.e.*, H1-8 mentioned in Table S1) are examined using two different degrees of confinement (*i.e.*, $\Gamma \in \{0.29, 0.40\}$). For all swimmers, $\Omega$ and $U$ show a linear relationship. The data points represent simulation results for different values of $\Gamma$. The dotted and dashed lines are the predictions of Eq. (3) for the different degrees of confinement using $b$=2.5. In all cases, Eq. (3) accurately fits the simulations' data points for $\Gamma > 0$. The loosely dashed and solid lines are the predictions of Eqs. (1) and (2) for the propulsion speed of the unconfined swimmer, respectively. For all swimmers, the propulsion speeds calculated using Eq. (2) are noticeably closer to the simulated propulsion speeds for $\Gamma = 0$ than those of Eq. (1).

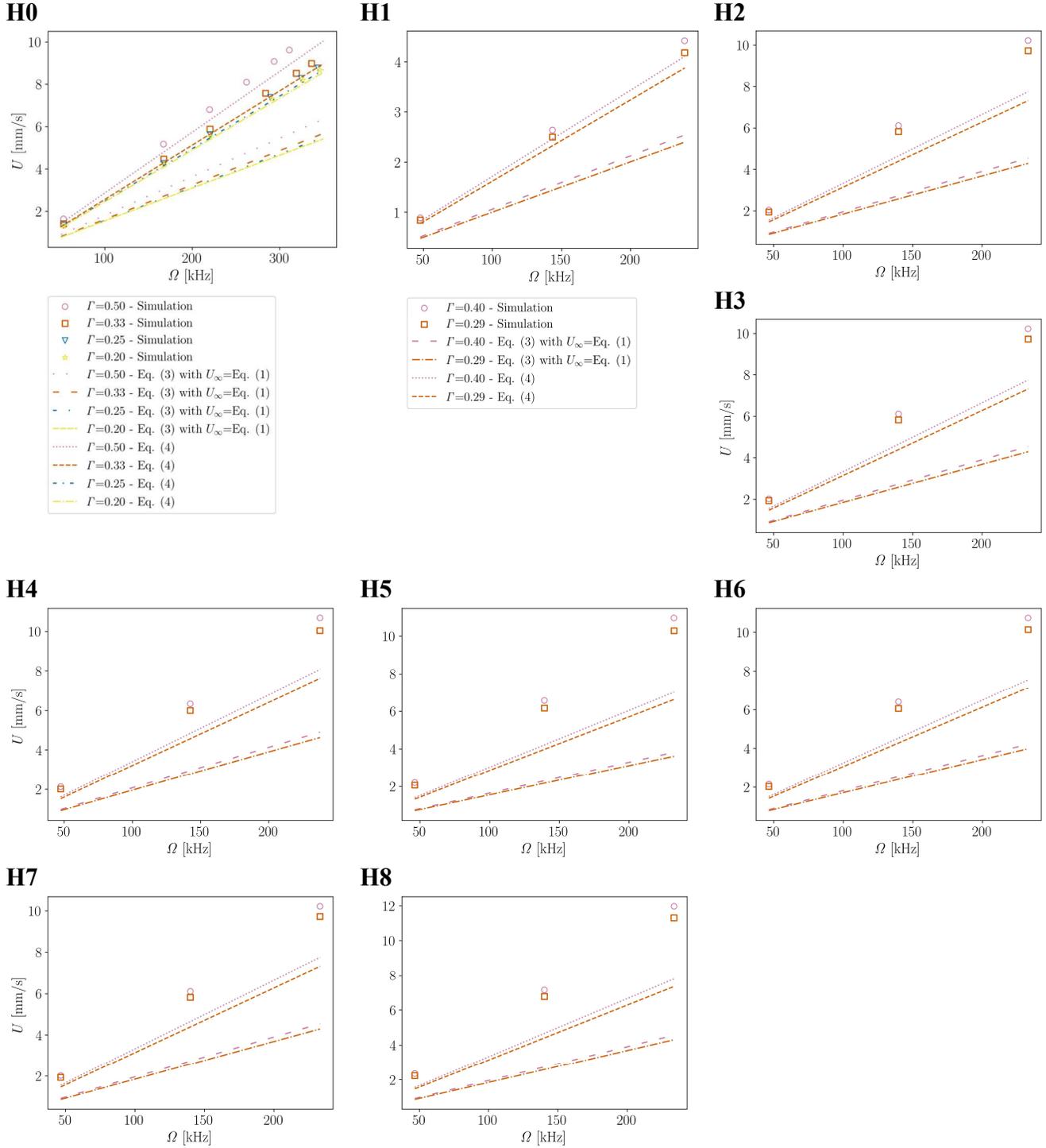

**Fig. S3. Purely analytical prediction for the confined propulsion speed of various helical nanoswimmers Hx.** Replacing the computationally estimated $U_\infty$ in Eq. (3) with the new analytical expression for the speed of an unconfined helix, *i.e.*, Eq. (2), results in a purely analytical expression, *i.e.*, Eq. (4), for the estimation of the confined swimming speed of helical nanoswimmers. The predictions from Eq. (4) for all swimmers listed in Table S1 are compared to the case in which the preexisting relation for the speed of an unconfined helix, *i.e.*, Eq. (1), is used for $U_\infty$ in Eq. (3). The results of Eq. (4) are significantly closer

to the simulation data points, providing reasonable estimates for the host environment-specific swimming speed of helical nanoswimmers without resorting to any expensive computations/experiments.

**Tables:**

**Table S1.** **Properties of the helical filaments (nanoswimmers) used in this work.** $\rho_w$ is the density of water. Previous analytical and computational studies, *e.g.*, Refs. (*42, 104, 105*), suggest that the optimal propellers have only about one helical turn. Note that the outer diameter of the swimmer is related to the major and minor radii via $D_0 = 2 \times (R_M + R_m)$.

| | Name | Major radius $R_M$ [nm] | Minor radius $R_m$ [nm] | Pitch angle $\alpha$ [deg] | Pitch length $P$ [nm] | Density $\rho$ [kg/m³] | Axial length $L_a$ [nm] |
|---|---|---|---|---|---|---|---|
| Fractional number of turns (*i.e.*, 1.25) (*42*) | H0 | 30 | 15 | 46.3 | 180.0 | $\rho_w$ | 255.0 |
| Varied major radius | H1 | 24 | 4.8 | 45 | 150.8 | $\rho_w$ | 160.4 |
| | H2 | 48 | 4.8 | 45 | 301.6 | $\rho_w$ | 311.2 |
| Varied minor radius | H3 | 48 | 4.8 | 45 | 301.6 | $\rho_w$ | 311.2 |
| | H4 | 48 | 8 | 45 | 301.6 | $\rho_w$ | 317.6 |
| Varied pitch angle | H5 | 48 | 4.8 | 35 | 430.7 | $\rho_w$ | 440.3 |
| | H6 | 48 | 4.8 | 40 | 359.4 | $\rho_w$ | 369.0 |
| Varied density | H7 | 48 | 4.8 | 45 | 301.6 | $\rho_w$ | 311.2 |
| | H8 | 48 | 4.8 | 45 | 301.6 | $20\rho_w$ | 311.2 |